\documentclass[aps,floatfix,preprint,showpacs,preprintnumbers,nofootinbib,superscriptaddress,natbib]{revtex4}

\usepackage{graphicx,float}
\usepackage{epsfig}		
\usepackage{dcolumn}
\usepackage{bm}
\usepackage{subfigure}

\usepackage[all]{xy}
\usepackage{amsmath,upgreek}
\usepackage{amssymb}

\usepackage{pdfpages}
\usepackage{color}
\usepackage{graphicx,epstopdf}
\usepackage[colorlinks,hyperindex]{hyperref}

\def\0{\mbox{\tiny $0$}}
\def\1{\mbox{\tiny $1$}}
\def\2{\mbox{\tiny $2$}}
\def\3{\mbox{\tiny $3$}}
\def\4{\mbox{\tiny $4$}}
\def\5{\mbox{\tiny $5$}}
\def\6{\mbox{\tiny $6$}}
\def\7{\mbox{\tiny $7$}}
\def\8{\mbox{\tiny $8$}}
\def\9{\mbox{\tiny $9$}}

\def\f14{\mbox{\tiny $\frac{1}{4}$}}

\begin{document}

\title{Bilayer graphene lattice-layer entanglement under non-Markovian phase noise}
\author{Victor A. S. V. Bittencourt}
\email{vbittencourt@df.ufscar.br}
\affiliation{Departamento de F\'{\i}sica, Universidade Federal de S\~ao Carlos, PO Box 676, 13565-905, S\~ao Carlos, SP, Brasil.}
\author{Massimo Blasone}
\email{blasone@sa.infn.it}
\affiliation{Dipartimento di Fisica, Universit\`a degli studi di Salerno, Via Giovanni Paolo II, 132 84084 Fisciano, Italy}
\altaffiliation{Also at: INFN Sezione di Napoli, Gruppo collegato di Salerno, Italy}
\author{Alex E. Bernardini}
\email{alexeb@ufscar.br}
\affiliation{Departamento de F\'isica e Astronomia, Faculdade de Ci\^{e}ncias daUniversidade do Porto, Rua do Campo Alegre 687, 4169-007, Porto, Portugal.}
\altaffiliation[On leave of absence from]{~Departamento de F\'{\i}sica, Universidade Federal de S\~ao Carlos, PO Box 676, 13565-905, S\~ao Carlos, SP, Brasil.}

\date{\today}
\renewcommand{\baselinestretch}{1.3}
\begin{abstract}
The evolution of single particle excitations of bilayer graphene under effects of non-Markovian noise is described with focus on the decoherence process of lattice-layer (LL) maximally entangled states.
Once that the noiseless dynamics of an arbitrary initial state is identified by the correspondence between the tight-binding Hamiltonian for the AB-stacked bilayer graphene and the Dirac equation -- which includes pseudovector- and tensor-like field interactions -- the noisy environment is described as random fluctuations on bias voltage and mass terms.
The inclusion of noisy dynamics reproduces the Ornstein-Uhlenbeck processes: a non-Markovian noise model with a well-defined Markovian limit. Considering that an initial amount of entanglement shall be dissipated by the noise, two profiles of dissipation are identified. On one hand, for eigenstates of the noiseless Hamiltonian, deaths and revivals of entanglement are identified along the oscillation pattern for long interaction periods. On the other hand, for departing LL Werner and Cat states, the entanglement is suppressed although, for both cases, some identified memory effects compete with the pure noise-induced decoherence in order to preserve the the overall profile of a given initial state.
\end{abstract}

\date{\today}
\maketitle

\section{Introduction}

Efforts to understand ground properties of graphene \cite{graph01, graph02, graph03, graph04, graphteo} have been in the streamline of both theoretical and experimental investigations on physics of nanostructures. The remarkable electronic properties of graphene result from a quite singular structure of its energy bands which exhibits a linear low energy profile driven by a massless Dirac-like equation \cite{graph01}. For example, under magnetic fields, the graphene conductance exhibits an anomalous behavior due to the formation of modified Landau levels \cite{graph05, graph06, graph07}. Likewise, in bilayer graphene, more suitable properties are driven by its weak interlayer coupling, which also depends on the particular double layer geometric arrangement \cite{graph08, graph010, graph011, graph012}. Differently from the single layer graphene, the energy bands of bilayer graphene display a hyperbolic structure near the corners of the first Brillouin zone. It reproduces the energy dispersion of free massive fermions and provides a subjacent correspondence with the Dirac equation structure which has been relevant into the investigation of relativistic-like effects (cf. {\em zitterbewegung} and the {\em Klein paradox} effects \cite{graph013, graph014, graph015}).

Apart from their electronic properties, graphene structures may also exhibit some quantum entanglement properties.
The study of entanglement in connection with quantum Hall effects \cite{CondMatter01, CondMatter02, CondMatter03} in graphene structures has shown a close relation between quantum correlations and their topological properties \cite{CondMatter04, CondMatter05}, even with some restrictions concerning the use of entanglement as a fingerprint for topological characterization \cite{CondMatterNew}. Graphene has also been tested as a quantum computing platform to implement quantum gates \cite{QuantumInfoGraph01, QuantumInfoGraph02, QuantumInfoGraph03, QuantumInfoGraph04} -- through, for instance, the spin-orbit coupling between a flying qubit and a graphene quantum dot used to engender either quantum logic operations \cite{QuantumInfoGraph01} or intervalley couplings \cite{ QuantumInfoGraph02, QuantumInfoGraph03}. 

Single-particle states of graphene also exhibit intrinsic entanglement according to their description through the Dirac equation \cite{PRB}. In such a framework, the solutions of Dirac equation are supported by a $SU(2)\otimes SU(2)$ group structure associated with two internal degrees of freedom (DoF's): the intrinsic parity and the spin.
The Dirac Hamiltonian is decomposed in terms of two-qubit operators which drives the dynamics of the Dirac bispinors identified in such a framework as two-qubit entangled states \cite{SU201, SU202}. In addition, the inclusion of global potentials into the Dirac dynamics modifies the spin-parity correlational content of Dirac equation solutions \cite{SU203}. A complete interacting Dirac Hamiltonian including external fields classified according to their invariance properties under Poincar\'e transformations \cite{Thaller} reads
\begin{eqnarray}
\label{E04T}
\hat{H} &=& A^0(x)\,\hat{I}_4+ \hat{\beta}[ m + \phi_S (x) ] + \hat{\bm{\alpha}} \cdot [\hat{\bm{p}} - \bm{A}(x)] + i \hat{\beta} \hat{\gamma}_5 \mu(x) - \hat{\gamma}_5 q(x) + \hat{\gamma}_5 \hat{\bm{\alpha}}\cdot\bm{W}(x) \nonumber \\
&+& i \hat{\bm{\gamma}} \cdot [ \chi_a \bm{B}(x) + \kappa_a\, \bm{E}(x) \,] + \hat{\gamma}_5 \hat{\bm{\gamma}}\cdot[\kappa_a\, \bm{B}(x) - \chi_a \bm{E}(x) \,],
\end{eqnarray}
with $\hbar = c = 1$, $\hat{\bm{\gamma}} = \hat{\beta} \hat{\bm{\alpha}}$, and $\hat{\gamma}_5 = -i \hat{\alpha}_x \hat{\alpha}_y \hat{\alpha}_z$, where
$\hat{\beta}$ and $\hat{\bm{\alpha}} = \{\hat{\alpha}_x, \, \hat{\alpha}_y, \, \hat{\alpha}_z \}$ are the Dirac matrices that satisfies the anti-commuting relations
$\{\hat{\alpha}_i, \hat{\alpha}_j\} = 2 \, \delta_{ij} \, \hat{I}_4$, and
$\{\hat{\alpha}_i, \hat{\beta}\} =0$, with $i,j = x,y,z$, and
$\hat{\beta}^2 = \hat{I}_4$ (where $\hat{I}_N$ denotes the $N$-dimensional identity operator). As a matter of simplicity, one considers the representation of Dirac matrices given by
\begin{equation}
\label{representation}
\hat{\bm{\alpha}} = \hat{\sigma}_x \otimes \hat{\bm{\sigma}} \equiv \left[ \begin{array}{rr} 0 & \hat{\bm{\sigma}} \\ \hat{\bm{\sigma}} & 0 \end{array}\right],
\qquad \mbox{and} \qquad
\hat{\beta} = \hat{\sigma}_z \otimes \hat{I}_2 \equiv \left[ \begin{array}{rr} \hat{I}_2 & 0 \\ 0 & - \hat{I}_2 \end{array} \right],
\end{equation}
where ${\bm{\sigma}}$ are the Pauli matrices, bold variables ``$\bm{a}$' denote vectors, with $a = \vert \bm{a} \vert = \sqrt{\bm{a} \cdot \bm{a}}$, and hats ``$~\hat{}~$'' denote operators. Apart from the free particle contribution, $\hat{\beta} m + \hat{\bm{\alpha}} \cdot \hat{\bm{p}}$, the Hamiltonian, Eq.~(\ref{E04T}), includes the interaction with an external vector field with time- and space-like components, $A^0(x)$ and $\bm{A} (x)$, and a non-minimal coupling to external magnetic and electric fields, $\bm{B}(x)$ and $\bm{E}(x)$ (through $\kappa_a$ and $\chi_a$, respectively). Interactions also involve an external pseudovector field, ($q(x)$,\,$\bm{W}(x)$), and both scalar and pseudoscalar fields, $\phi_S(x)$ and $\mu(x)$.

In the most stable configuration of the bilayer graphene, the AB (or Bernal) stacking, the tight-binding (TB) Hamiltonian governing low energy excitations can be written as a Dirac Hamiltonian including pseudovector and tensor external fields, such that the dynamics of single particle excitations of the system can be recovered through the bispinor solutions of the corresponding Dirac equation \cite{PRB}. The $SU(2) \otimes SU(2)$ entangled structure of the Dirac equation is thus translated into an intrinsic lattice-layer (LL) entanglement carried by single particle states. A complete description of LL entanglement then include effects of the on-site interactions associated with bias voltage and mass terms in the tight binding prescription \cite{PRB}. 

Departing from graphene structures preliminary described as closed quantum systems \cite{PRB}, the aim of our work is to compute the influence of a noisy dynamics on the intrinsic LL entanglement. The framework is driven by a non-Markovian noise model which posses a well-defined Markovian limit where classical random frequency fluctuations are modeled by the Ornstein-Uhlenbeck process \cite{OUProcess01, OUProcess02} which is, by the way, included into the dynamics driven by the tight binding Hamiltonian.
It is assumed that the lattice and the layer DoF's are separately affected by the environment, through interaction terms representing random fluctuations of the bias voltage and the mass terms of the TB model. The noisy evolution is included via Kraus operators, and the complete dynamics of an arbitrary initial state as well as the time evolution of its quantum entanglement are obtained.

The paper is organized as follows. In Sec. II, a brief review of the TB model for the AB-stacked bilayer graphene along with its connection to the Dirac Hamiltonian is introduced. In Sec. III, the time evolution of an arbitrary initial state under the noiseless dynamics is recovered, and the dynamics of maximally entangled states are described. Sec. IV introduces the classical noise model via Kraus operators and the dynamics of LL states under the noisy dynamics is built. The effects of the non-Markovian fluctuations on Hamiltonian eigenstates and on LL Cat and Werner states are all obtained. Final conclusions and next-step perspectives are drawn in Sec. V. 

\section{Tight-binding Hamiltonian and its relation to the Dirac equation}

One effective description of bilayer graphene, often considered for describing electronic and optical properties, is the TB approach given by the Hamiltonian
\begin{eqnarray}
\label{tightbinding}
\hat{\mathcal{H}}_{AB} = &-& t \displaystyle \sum_{\bm{k}} \left[ \, \Gamma(\bm{k}) \hat{a}_{1\bm{k}} ^\dagger \hat{b}_{1\bm{k}} + \Gamma(\bm{k}) \hat{a}_{2 \bm{k}}^\dagger \hat{b}_{2 \bm{k}} + \mbox{h.c.} \,\right] \nonumber \\
 &+& t_{\bot} \displaystyle \sum_{\bm{k}} \left[ \, \hat{b}_{1 \bm{k}}^\dagger \hat{a}_{2 \bm{k}} + \hat{a}_{2 \bm{k}}^\dagger \hat{b}_{1 \bm{k}} \, \right] - t_3 \displaystyle \sum_{\bm{k}} \left[\, \Gamma(\bm{k}) \hat{b}_{2 \bm{k}}^\dagger \hat{a}_{1 \bm{k}} + \Gamma^* (\bm{k}) \hat{a}_{1 \bm{k}}^\dagger \hat{b}_{2 \bm{k}} \, \right] \nonumber \\
 &+&t_4 \displaystyle \sum_{\bm{k}} \left[ \, \Gamma(\bm{k})(\hat{a}_{1 \bm{k}}^\dagger \hat{a}_{2 \bm{k}} + \hat{b}_{1 \bm{k}}^\dagger \hat{b}_{2 \bm{k}}) + \mbox{h.c.} \, \right],
\end{eqnarray}
 where $\hat{\alpha}_{i\, \bm{k}}^\dagger$ is the creation operator for an excitation on the $\alpha$ lattice in the $i$-th layer, with the wave vector $\bm{k}$, and $\Gamma(\bm{k}) = \sum_{j=1}^3 e^{i \bm{k} \cdot \bm{\delta}_j}$ is given in terms of the vectors
\begin{eqnarray}
\bm{\delta}_{1,2} = \left(- \frac{a}{2}, \, \pm \frac{a \sqrt{3}}{2} \right), \hspace{0.6 cm} \bm{\delta}_3 = (a ,\, 0),
\end{eqnarray}
connecting a given site to its nearest-neighbor. The hopping amplitudes $t$, $t_3$, $t_\bot$ and $t_4$ are schematically depicted in the Appendix, and their experimental values, obtained via infrared spectroscopy \cite{ExperimentalParameters01}, are given by
\begin{eqnarray}
\label{experimentalpar}
t &=& 3.16 \, \pm \, 0.03 \, \, \mbox{eV}, \hspace{0.5 cm} t_\bot =  0.381 \, \pm \, 0.003 \, \, \mbox{eV}, \nonumber \\
t_3 &=& 0.38 \, \pm 0.06 \, \, \mbox{eV},  \hspace{0.6 cm} t_4 = 0.14 \, \pm \, 0.03 \, \, \mbox{eV}, 
\end{eqnarray}
which, a part for the hopping $t$, are approximately the same values obtained via DFT calculations \cite{ExperimentalParameters02}.

To sustain the analytical approach, one sets $t_4 = 0$, and the TB Hamiltonian in $\bm{k}$ space is written in the basis $\{ \vert A1 (\bm{k}) \rangle, \vert B1 (\bm{k}) \rangle, \vert A2 (\bm{k}) \rangle, \vert B2 (\bm{k}) \rangle \}$ ($\vert \alpha _i (\bm{k}) \rangle = \hat{\alpha}_{i \bm{k}} ^\dagger \vert 0 \rangle$) as
\begin{equation}
\label{ABHamiltonian}
\hat{\mathcal{H}}_{AB} =\left[ \, \begin{array}{cccc}
0 & - t \Gamma(\bm{k}) & 0 & - t_3 \Gamma^*(\bm{k})\\
- t \Gamma^*(\bm{k}) & 0 & t_\bot & 0\\
0 & t_\bot & 0& -t\Gamma (\bm{k})\\
- t_3 \Gamma(\bm{k}) & 0 & - t \Gamma^*(\bm{k}) & 0
\end{array} \right].
\end{equation}
One may also consider two additional {\em on site} interactions which open an energy gap between the valence and the conduction bands, the mass term and the bias-voltage, given respectively by \cite{graph03, Predin}
\begin{equation}
\label{massHamiltonian}
\hat{\mathcal{H}}_{m}= \mbox{diag}\{ m, \, -m, \, m, \, -m\},
\end{equation}
\begin{equation}
\label{biasvoltageHamiltonian}
\hat{\mathcal{H}}_{\Lambda}= \mbox{diag} \left\{ \frac{\Lambda}{2}, \, \frac{\Lambda}{2}, \, - \frac{\Lambda}{2}, \, - \frac{\Lambda}{2} \right\},
\end{equation}
as to have the total Hamiltonian in $\bm{k}$ space \cite{PRB} written as
\begin{eqnarray}
\label{Htotal}
\hat{\mathcal{H}} &=& \hat{\mathcal{H}}_{AB} + \hat{\mathcal{H}}_m + \hat{\mathcal{H}}_{\Lambda} = \left[ \, \begin{array}{cccc}
m+ \frac{\Lambda}{2} & - t \Gamma(\bm{k}) & 0 & - t_3 \Gamma^*(\bm{k})\\
- t \Gamma^*(\bm{k}) & -m + \frac{\Lambda}{2} & t_\bot & 0\\
0 & t_\bot & m - \frac{\Lambda}{2} & -t\Gamma (\bm{k})\\
- t_3 \Gamma(\bm{k}) & 0 & - t \Gamma^*(\bm{k}) & -m - \frac{\Lambda}{2}
\end{array} \right],
\end{eqnarray}

which can be rewritten in the form of the modified Dirac Hamiltonian,
\begin{equation}
\label{DiracModified}
\hat{\mathcal{H}} = \bm{p} \cdot \hat{\bm{\alpha}} + M \hat{\beta} + \bm{W} \cdot \, \hat{\gamma}_5 \hat{\bm{\alpha}} + i \bm{\mathcal{E}} \cdot \hat{\bm{\gamma}}.
\end{equation}
In comparison with Eq.~(\ref{E04T}), the Dirac form involves the usual free particle term, $ \bm{p} \cdot \hat{\bm{\alpha}} + M \hat{\beta}$, and it includes pseudovector and pseudotensor contributions, $ \bm{W} \cdot \, \hat{\gamma}_5 \hat{\bm{\alpha}}$ and $i \bm{\mathcal{E}} \cdot \hat{\bm{\gamma}}$.
If one notices that the total TB Hamiltonian, Eq.~(\ref{Htotal}), can be decomposed in terms of the Dirac matrices as
\begin{eqnarray}
\hat{\mathcal{H}} &=&
\frac{t_\bot}{2}\left( \hat{\alpha}_x - i \hat{\gamma}_y\right)
- t \left\{\mbox{Re}[\Gamma (\bm{k})] \hat{\gamma}_5 \hat{\alpha}_x -\mbox{Im}[\Gamma(\bm{k})] \hat{\gamma}_5 \hat{\alpha}_y\right\}\nonumber\\
&&\qquad\qquad -\frac{t_3}{2}\left\{ \mbox{Re}[\Gamma (\bm{k})](\hat{\alpha}_x + i\,\hat{\gamma}_y) + \mbox{Im}[\Gamma(\bm{k})]( \hat{\alpha}_y - i\, \hat{\gamma}_x)\right\}
+ m \hat{\gamma}_5 \hat{\alpha}_z
+\frac{\Lambda}{2}\hat{\beta},
\end{eqnarray}
one sets the following correspondence between graphene and Dirac parameters
\begin{eqnarray}
\label{relations}
\bm{p} &\leftrightarrow& \frac{t_\bot - t_3 \, \mbox{Re}[\Gamma (\bm{k})]}{2} \bm{i} - \frac{t_3 \, \mbox{Im}[\Gamma(\bm{k})]}{2} \bm{j}, \hspace{0.5 cm } M \leftrightarrow \frac{\Lambda}{2} ,\nonumber \\
\bm{W} &\leftrightarrow& - t \, \mbox{Re}[\Gamma (\bm{k})] \bm{i} + t \, \mbox{Im}[\Gamma(\bm{k})] \bm{j} + m \bm{l}, \hspace{0.5 cm} \bm{\mathcal{E}} \leftrightarrow \frac{ t_3 \, \mbox{Im}[\Gamma(\bm{k})]}{2} \bm{i} - \frac{t_\bot + t_3 \, \mbox{Re}[\Gamma(\bm{k})]}{2} \bm{j},
\end{eqnarray}
where $\{\bm{i}, \bm{j}, \bm{l}\}$ are unitary vectors. The relation between the Hamiltonians (\ref{Htotal}) and (\ref{DiracModified}) can be interpreted as a simulation of the Dirac equation by the TB model.
In this framework, the eigenstates of the modified Dirac Hamiltonian, $\vert \, \psi_{n\, s} \, \rangle$ ($n,s = \{0,1\}$), are written as \cite{PRB}
\begin{equation}
\label{eigenstates}
\vert \psi_{n,s} (\bm{k}) \rangle \equiv M^{A1}_{n,s}\, \,\vert A1 (\bm{k}) \rangle + M^{B1}_{n,s} \, \,\vert B1 (\bm{k}) \rangle + M^{A2}_{n,s} \,\,\vert A2 (\bm{k}) \rangle+ M^{B2}_{n,s} \, \,\vert B2 (\bm{k}) \rangle.
\end{equation}

Most importantly, the modified Dirac Hamiltonian (\ref{DiracModified}) possesses some algebraic properties by means of which the eigenstates can be straightforwardly calculated \cite{SU203} and, due to the relation with the TB Hamiltonian, the complete set of eigenstates and eigenvalues can be recovered for graphene one-particle excitations \cite{PRB}. The calculation procedure \cite{SU203} is supported by the properties of the traceless gamma matrices and it involves writing the a squared Hamiltonian operator as
\begin{eqnarray}
\label{prop01}
\hat{\mathcal{H}}^2 &=& g_1 \hat{I}_4 + 2 \hat{\mathcal{O}},
\end{eqnarray}
which, from Eq.~(\ref{DiracModified}), involves the traceless operator
\begin{equation}
\hat{\mathcal{O}} = (\bm{p} \cdot \bm{W}) \hat{\gamma}_5 + i (\bm{W}\cdot \bm{\mathcal{E}}) \hat{\beta} \hat{\gamma}_5 - [\, M \bm{W} + (\bm{p} \times \bm{\mathcal{E}}) \,] \cdot \hat{\gamma}_5 \hat{\bm{\gamma}},
\end{equation}
that returns 
\begin{equation}
\label{prop02}
\hat{\mathcal{O}}^2 = \frac{1}{4} \bigg{(} \hat{\mathcal{H}}^2 - g_1 \hat{I}\bigg{)}^2 = g_2 \hat{I},
\end{equation}
in terms of the auxiliary coefficients
\begin{eqnarray}
\label{gs}
g_1 &=& \frac{1}{4}\mbox{Tr}[\hat{\mathcal{H}}^2] = p^2 + M^2 + W^2 + \mathcal{E}^2. \nonumber \\
g_2 &=& \frac{1}{16}\mbox{Tr}\left[\,( \hat{\mathcal{H}}^2 - \frac{1}{4}\mbox{Tr}[\hat{\mathcal{H}}^2] )^2 \right] = \nonumber \\
&=& M^2 W^2 + 2 M \bm{W} \cdot(\bm{p} \times \bm{\mathcal{E}}) + \vert \bm{p} \times \bm{\mathcal{E}} \vert^2 + (\bm{p} \cdot \bm{W})^2 + (\bm{W} \cdot \bm{\mathcal{E}})^2.
\end{eqnarray}
The eigenstate density matrices $\rho_{n,s} = \vert \psi_{n,s} \rangle \langle \psi_{n,s} \vert$ of the Hamiltonian satisfying the relations (\ref{prop01})-(\ref{prop02}) are given \cite{SU203}
\begin{equation}
\label{ansatz}
\rho_{n,s} = \frac{1}{4}\left[\hat{I}_4 + \frac{(-1)^n}{\vert \lambda_{n,s} \vert} \, \hat{\mathcal{H}} \right] \left[ \hat{I}_4 + \frac{(-1)^s}{\sqrt{g_2}} \, \hat{\mathcal{O}} \right],
\end{equation}
which are stationary states of the corresponding Liouville equation $[\hat{\mathcal{H}}, \rho_{n,s}] = 0$. The eigenenergies, $\lambda_{n,s}$, evaluated by the averaged value of the Hamiltonian read
\begin{equation}
\label{ener}
\lambda_{n,s} = \mbox{Tr}[\hat{\mathcal{H}} \rho_{n,s}]= (-1)^n\sqrt{g_1 + 2 (-1)^s \sqrt{g_2}}.
\end{equation}

The single particle energy spectrum of the bilayer graphene in $\bm{k}$ space can be recovered by substituting the relation (\ref{relations}) into (\ref{gs}) and (\ref{ener}) so to result into
\begin{eqnarray}
\label{eigenvalues}
\lambda_{n,s}(\bm{k}) &=& (-1)^n \bigg[ \frac{1}{2} \bigg( 2 t^2 \vert \Gamma(\bm{k}) \vert^2 + t_\bot^2 + t_3 \vert \Gamma(\bm{k}) \vert^2 + 2 m^2 + \frac{\Lambda^2}{2} \nonumber \\
 &&\qquad\qquad\qquad+ (-1)^s [4 t^2\, \vert \Gamma(\bm{k}) \vert^2 (\,t_\bot^2 + \Lambda^2+ t_3^2 \vert \Gamma(\bm{k}) \vert^2 - 2 t_\bot t_3 \cos(3 \phi(\bm{k}))) \nonumber \\ &&\qquad\qquad\qquad\qquad\qquad\qquad + ( t_3 \vert \Gamma(\bm{k}) \vert^2 - t_\bot^2 + 2 m \Lambda )^2 ]^{1/2} \, \bigg) \bigg]^{1/2},
\end{eqnarray}
where $\Gamma(\bm{k}) = \vert \Gamma(\bm{k}) \vert e^{i \phi(\bm{k})}$. The hyperbolic dispersion relation defined by the $\lambda_{n,s}$ is composed by two energy branches (associated to $s=0$ and $s=1$) and two energy bands (associated to $n=0$ and $n=1$). The energy bands exhibit extremum points for specific values of the wave vector $\bm{k}$. In particular, two extrema occur when $\Gamma(\bm{k}) = 0$, which corresponds to two inequivalent Dirac points
\begin{equation}
\label{diracpoints}
\bm{K}_\pm = \frac{2 \pi}{3 \sqrt{3} a}(\sqrt{3}, \pm 1).
\end{equation}

\section{Lattice-layer entanglement and noiseless evolution of Cat and Werner states}

As to evince the correlation properties driven by the modified Dirac Hamiltonian, one rewrites Eq.~(\ref{DiracModified}) in terms of tensor products of Pauli matrices
\begin{equation}
\hat{\mathcal{H}} = \bm{p} \cdot (\hat{\sigma}_x^{(1)} \otimes \hat{\bm{\sigma}}^{(2)}) + M (\hat{\sigma}_z^{(1)} \otimes \hat{I}^{(2)}) + \bm{W} \cdot (\hat{I}^{(1)} \otimes \hat{\bm{\sigma}}^{(2)}) - \bm{\mathcal{E}} \cdot(\hat{\sigma}_y^{(1)} \otimes \hat{\bm{\sigma}}^{(2)}), 
\end{equation}
thus interpreting the dynamics driven by such Hamiltonian as describing the evolution of two discrete DoF's associated to the labels (1) and (2). The states evolving under such dynamics describe a system $\mathcal{S}$ composed by two subsystems, $\mathcal{S}_1$ (associated with the spin DoF) and $\mathcal{S}_2$ (associated with the intrinsic parity DoF) supported by a Hilbert space $H = H_1 \otimes H_2$ with $\mbox{dim} H_1 = \mbox{dim} H_2 = 2$. Moreover, the corresponding eigenstates (\ref{DiracModified}) are bipartite parity-spin entangled states \cite{SU201, SU202}, and this $SU(2)\otimes SU(2)$ structure sets the condition for computing entanglement quantifiers.
As preliminarily investigated in various scenarios \cite{SU203, Barrier, PRA}, a bipartite state described by a density operator $\rho \in H_1 \otimes H_2$ is separable if
\begin{equation}
\rho = \displaystyle \sum_i w_i \hat{\tau}_i^{(1)} \otimes \hat{\tau}_i^{(2)},
\end{equation}
where $\hat{\tau}_i^{(j)} \in H_j$, $w_i >0$ and $\sum_i w_i = 1$. The separability concept can be translated in terms of the Peres criterion, which establishes that for a state to be separable, all eigenvalues of its partial transpose density matrix must be positive \cite{Entanglement01}. It fits the entanglement measure criterium which shall be persecuted along this paper.
According to the Peres criterion, the entangled measure of a two-qubit state $\rho$ -- the so-called negativity -- is defined as \cite{Negativity}
\begin{equation}
\label{negativity}
\mathcal{N}[\rho] = \vert \vert \, \rho_1 ^T \, \vert \vert -1 = \displaystyle \sum_i \vert \mu_i \vert -1,
\end{equation}
where $\vert \vert \, \rho_1 ^T \, \vert \vert = \displaystyle \sum_i \vert \mu_i \vert $ is the trace norm of the matrix $\rho_1$, with eigenvalues $\mu_i$, obtained through the partial transposition of the original density matrix $\rho$ with respect to the subsystem $1$. With respect to a fixed basis on the composite Hilbert space $\{\vert \mu_i \rangle \otimes \vert \nu_j \rangle \}$ (with $\vert \mu_i \rangle \in H_{1}$ and $\vert \nu_i \rangle \in H_{2}$), the matrix elements of the partial transpose with respect to the first subsystem $\rho_1 ^T$ are given by
\begin{equation}
\langle \mu_i \vert \otimes \langle \nu_j \vert \rho_1 ^T \vert \mu_k \rangle \otimes \vert \nu_l \rangle = \langle \mu_k \vert \otimes \langle \nu_j \vert \, \rho \, \vert \mu_i \rangle \otimes \vert \nu_l \rangle.
\end{equation}

Turning back to the one-to-one correspondence between the bilayer graphene Hamiltonian (\ref{Htotal}) and the modified Dirac Hamiltonian (\ref{DiracModified}), one can identify the two DoF's intrinsic to bilayer graphene dynamics (cf. Eq.~(\ref{Htotal})) as lattice ($A$ or $B$) and layer ($1$ or $2$) \cite{PRB}, such that the intrinsic spin-parity entanglement of Dirac bispinors corresponds to the LL entanglement. One particle states of the bilayer graphene can thus be interpreted as two-qubit states, and from now on the quibit assignment shall be given by
\begin{eqnarray}
\label{assi}
\vert A1 \rangle &\equiv& \vert 00 \rangle, \hspace{0.5 cm} \vert B1 \rangle \equiv \vert 01 \rangle, \nonumber \\
\vert A2 \rangle &\equiv& \vert 10 \rangle, \hspace{0.5 cm} \vert B2 \rangle \equiv \vert 11 \rangle.
\end{eqnarray}
In particular, the eigenstates as given by Eq.~(\ref{eigenstates}) are, in general, LL entangled.
In the summary of entaglement properties investigated in Ref.~\cite{PRB}, the absence of the gapping terms, (\ref{massHamiltonian}) and (\ref{biasvoltageHamiltonian}), leads to eigenstates (with wave vectors near to the Dirac points) with high degree of entanglement \cite{PRB}. In particular, it has been shown that the bias voltage term (\ref{biasvoltageHamiltonian}) spreads entanglement around the Dirac points, while the mass term (\ref{massHamiltonian}) has an overal effect of destroying LL entanglement of the eigenstates.
Therefore, to avoid misconceptions relative to the inclusion of noise effects, from now on one sets $m=0$ since its contribution has already been investigated in Ref.~\cite{PRB}.

Given a generic one-particle state of the graphene bilayer Hamiltonian represented by its density matrix $\rho$, through the qubit assignment (\ref{assi}) it is possible to evaluate the LL entanglement with the negativity (\ref{negativity}). Moreover, the completeness relation satisfied by the density matrix of the eigenstates $\sum_{\{n,s\}} \rho_{n,s} = \hat{I}$ allows the reconstruction of the temporal evolution of any initial state $\rho(0)$ through
\begin{equation}
\label{timeevo}
\rho (\tau) = e^{- i \hat{\mathcal{H}} \tau} \rho(0) e^{ i \hat{\mathcal{H}} \tau} = \displaystyle \sum^1_{n,s=0}\sum^1_{m,l=0} e^{- i (\lambda_{n,s} - \lambda_{m,l}) \tau}\, \varrho_{n,s} \, \rho (0) \, \varrho_{m,l}.
\end{equation}
Given the dynamics obtained through the above equation, one can evaluate the mean value of any observable $\hat{A}$ through $\langle \hat{A} \rangle (\tau) = \mbox{Tr}[\hat{A} \rho(\tau)]$. In particular, the survival probability, i.e. the probability of measuring $\rho(\tau)$ in its initial configuration, is evaluated by
\begin{eqnarray}
\label{survivprob}
\mathcal{P}_{\rho(0)} (\tau) &=& \mbox{Tr}[\rho(0) \rho(\tau)] = \displaystyle \sum_{n,s =0}^1 \, \sum_{m,l = 0}^1 e^{- i (\lambda_{n,s} - \lambda_{m,l}) \tau} \mbox{Tr}[\rho(0) \rho_{n,s} \rho(0) \rho_{m,l}].
\end{eqnarray}
In the above framework it is possible to reconstruct the dynamical behavior of any initial one-particle state under the dynamics specified by the Hamiltonian (\ref{Htotal}), as the eigenstates are in terms of the Dirac eigenstates.

Maximally entangled states LL states can be constructed as $\rho_C(\tau = 0) = \vert \psi_C \rangle \langle \psi_C \vert$ (the Cat state) and $\rho_W(\tau = 0) = \vert \psi_W \rangle \langle \psi_W \vert$ (the Werner state):
\begin{eqnarray}
\label{cwstates}
\vert \psi_C \rangle &=& \frac{a^\dagger_1 (\bm{k}) + b^\dagger_2 (\bm{k})}{\sqrt{2}} \vert 0 \rangle = \frac{\vert A1 (\bm{k}) \rangle + \vert B2 (\bm{k}) \rangle}{\sqrt{2}}, \nonumber \\
 \vert \psi_W \rangle &=& \frac{a^\dagger_2 (\bm{k}) + b^\dagger_1 (\bm{k})}{\sqrt{2}} \vert 0 \rangle = \frac{\vert A2 (\bm{k}) \rangle + \vert B1 (\bm{k}) \rangle}{\sqrt{2}}.
\end{eqnarray}
The time evolution of $\rho_C (\tau = 0)$ and $\rho_W (\tau = 0)$ are obtained through Eq.~(\ref{timeevo}), and the corresponding survival probabilities $\mathcal{P}_C (\tau)$ and $\mathcal{P}_W(\tau)$ are recovered through Eq.~(\ref{survivprob}). For a given wave vector $\bm{k}$, $\rho_C (\tau = 0)$ and $\rho_W (\tau = 0)$, the LL entanglement returns the maximal value $\mathcal{N}[\rho_{C (W)} (0)] = 1$. Once the time evolution of the Dirac states is specified, the temporal evolution of entanglement is straightforwardly obtained in terms of the associated  negativity.

For Werner and Cat states, with wave vectors in the corner of the first Brillouin zone, the analysis of the noiseless temporal evolution can be simplified. The Hamiltonian (\ref{Htotal}) for $\bm{k} = \bm{K}_+$ (and for $m = 0$) reads
\begin{equation}
\label{simplifiedHamiltonian}
\hat{H}_{\bm{k} = \bm{K}_+}=\frac{1}{2}\left[
\begin{array}{rrrr}
+\Lambda & 0 & 0 & 0 \\
0 & +\Lambda & 2 t_{\bot} & 0 \\
0 & 2 t_{\bot} & -\Lambda & 0 \\
0 & 0 & 0 & -\Lambda \\
\end{array}
\right],
\end{equation}
a matrix form Hamiltonian composed by two blocks respectively space spanned by $\{\vert 00 \rangle, \vert 11 \rangle\}$ and by $\{\vert 01 \rangle, \vert 10 \rangle\}$. For $\bm{k} = \bm{K}_+$, the \textit{ansatz} Eq.~(\ref{ansatz}) returns the four eigenstates given explicitly by
\footnotesize
\begin{eqnarray}
\label{EigenstatesCorner}
\rho_{n, 0}=\left[
\begin{array}{cccc}
\vspace{0.2 cm} 0 & 0 & 0 & 0 \\ \vspace{0.2 cm}
0 & \frac{(-1)^n \Lambda + \sqrt{4 t_{\bot}^2 + \Lambda^2}}{2\, \sqrt{4 t_{\bot}^2 + \Lambda^2} } & (-1)^n \, \frac{t_\bot}{\sqrt{4 t_{\bot}^2 + \Lambda^2} } & 0 \\ \vspace{0.2 cm}
0 &(-1)^n \frac{t_\bot}{\sqrt{4 t_{\bot}^2 + \Lambda^2} } & \frac{ (-1)^{n+1}  \, \Lambda + \sqrt{4 t_{\bot}^2 + \Lambda^2}}{2 \,\sqrt{4 t_{\bot}^2 + \Lambda^2} } & 0 \\
0 & 0 & 0 & 0 \\
\end{array}
\right], \hspace{0.5 cm} \rho_{n,1} = \left[
\begin{array}{cccc}
\vspace{0.2 cm} \delta_{n,0} & 0 & 0 & 0 \\ \vspace{0.2 cm}
0 &0 & 0 & 0 \\ \vspace{0.2 cm}
0 &0 &0 & 0 \\
0 & 0 & 0 & \delta_{n,1} \\
\end{array}
\right],
\end{eqnarray}
\normalsize
as to give $\rho_{n,0}$ as a linear combination of $\vert 0 1 \rangle$ and $\vert 1 0 \rangle$, and $\rho_{n,1}$ as a linear combination of $\vert 0 0 \rangle$ and $\vert 1 1 \rangle$. Therefore, in this case, the Werner state, which is a linear combination of the eigenstates described by $\rho_{0,0}$ and $\rho_{1,0}$, and the Cat state, which is a linear combination of the eigenstates described by $\rho_{0,1}$ and $\rho_{1,1}$, both have their temporal evolution simplified.

Due to the block structure of the Hamiltonian Eq.~(\ref{simplifiedHamiltonian}), $\rho_C(\tau)$ does not overlaps with $\rho_{n,0}$, while $\rho_W(\tau)$ does not overlaps with $\rho_{n,1}$. Through the Eqs.~(\ref{timeevo}) and (\ref{EigenstatesCorner}) one has, for the Cat state,
\begin{equation}
\rho_C(\tau) = \frac{1}{2}\left[
\begin{array}{cccc}
1 & 0 & 0 & e^{-i \Lambda \, \tau} \\
0 & 0 & 0 & 0 \\
0 & 0 & 0 & 0 \\
 e^{i \Lambda \, \tau} & 0 & 0 & 1 \\
\end{array}
\right],
\end{equation}
and, for the Werner state,
\begin{equation}
\rho_W(\tau) =\frac{1}{2} \left[
\begin{array}{cccc}
0 & 0 & 0 & 0 \\
0 & 1 + \mathcal{A}(\tau) & \mathcal{B}(\tau) & 0 \\
0 & \mathcal{B}^*(\tau) & 1 + \mathcal{A}(\tau) & 0 \\
0 & 0 & 0 & 0
\end{array} \right],
\end{equation}
with
\begin{eqnarray}
\mathcal{A}&=& \frac{2\Lambda   \, \left(\, t_\bot-\cos \bigg{(} \tau \sqrt{\Lambda^2+4 t_\bot^2} \bigg{)} \,\right)}{\Lambda^2+4 t_\bot^2},  \\
\mathcal{B} &=& \frac{4 t_\bot}{\Lambda^2+4t_\bot^2} \left[ \, t_{\bot} + \frac{\Lambda}{4} \left( \,  \Lambda   \cos \left(\tau\sqrt{4 t_\bot^2 + \Lambda^2} \right)  - i  \sqrt{\Lambda^2+4 t_\bot^2} \sin \left( \tau\sqrt{4 t_\bot^2 + \Lambda^2}\right) \right) \right]. \quad\,
\end{eqnarray}
The corresponding expressions for the survival probabilities are then given by
\begin{eqnarray}
\label{SurvivProbs}
\mathcal{P}_C &=& \mbox[Tr][ \rho_C (\tau) \, \rho_{C}(\tau = 0)] = \cos ^2\left(\frac{1}{2} \Lambda \, \tau \right), \nonumber \\
\mathcal{P}_W &=& \mbox[Tr][ \rho_W (\tau) \, \rho_{W}(\tau = 0)] = \frac{1}{2 \left(\Lambda^2+4 t_\bot^2 \right)} \left[\, 8 t_\bot^2 + \Lambda^2 \left( 1 + \cos {(\tau\sqrt{4 t_\bot^2 + \Lambda^2} )} \, \right) \right],
\end{eqnarray}
respectively for Cat and Werner states, and the temporal evolution of the quantum entanglement results into the follow expressions for the negativity, 
\begin{eqnarray}
\label{Negs}
\mathcal{N}[\rho_C(\tau)] &=& 1, \nonumber \\
\mathcal{N}[\rho_W(\tau)] &=& \frac{1}{\Lambda^2 + 4 t_\bot^2} \Big[ 16 t_\bot^4 + \Lambda^4 + \nonumber \\
&&
\qquad+ 4 \Lambda^2 t_\bot^2 \left(2 \cos{(\tau\sqrt{4 t_\bot^2 + \Lambda^2})} + \sin^2{(\tau\sqrt{4 t_\bot^2 + \Lambda^2})}  \right) \Big]^{1/2}.
\end{eqnarray}

\begin{figure}[H]
\centering
\includegraphics[width= 8 cm]{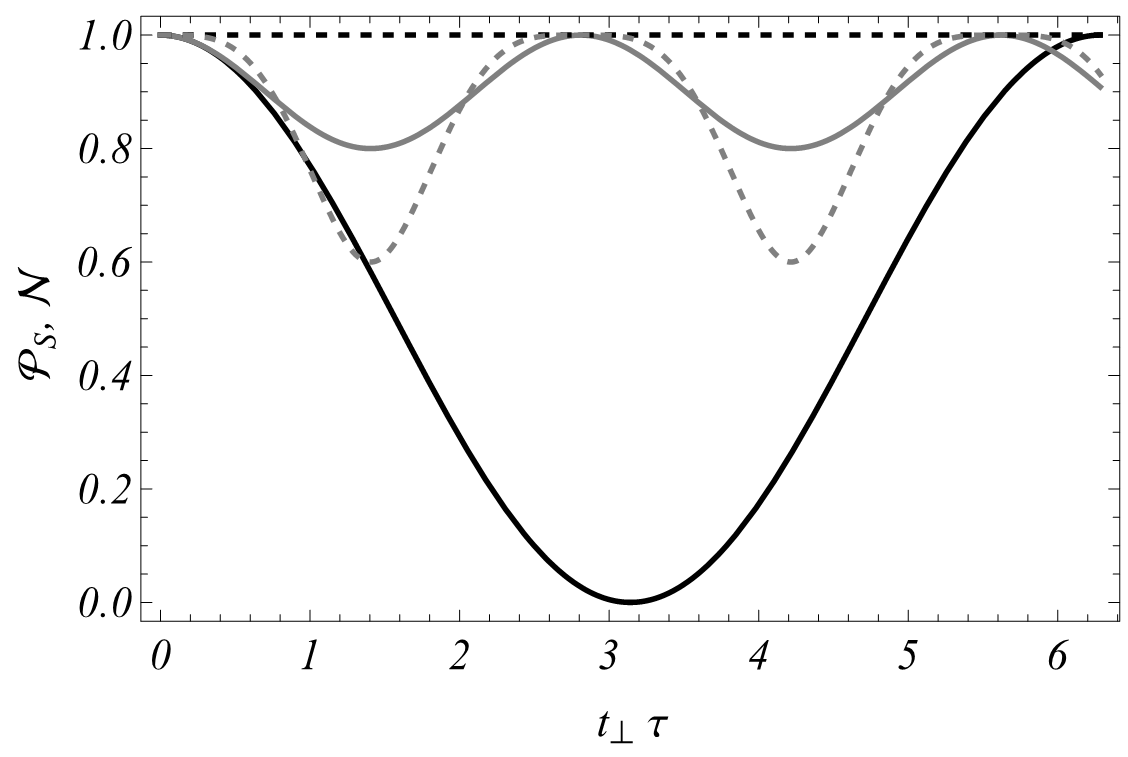}
\renewcommand{\baselinestretch}{1.0}
\caption{Survival probabilities (continuous lines) and entanglement (dashed lines) as function of the dimensionless parameter $t_\bot \tau$ for the Cat state, $\rho_C(\tau=0)$ (black lines), and the Werner state, $\rho_W (\tau = 0)$ (gray lines), where it has been adopted $\Lambda/t_\bot = 1$.  Both survival probabilities oscillate in time due to overlapping of the initial state with different eigenstates of the Hamiltonian, and the oscillations exhibit a well-defined periodicities given by (\ref{periodicites}). The initial Cat state retains its amount of LL entanglement during the time evolution, while the negativity of the initial Werner state oscillates with the same frequency of its survival probability.}
\label{FigNoiselessSurvivor}
\end{figure}

Fig.~\ref{FigNoiselessSurvivor} depicts the survival probabilities (continuous lines) and the negativity (dashed lines) for initial Cat (black lines) and Werner (gray lines) states with wave vectors in the corner of the first Brillouin zone $\bm{K}_+$ (\ref{diracpoints}),  given explicitly by Eqs.~(\ref{SurvivProbs})-(\ref{Negs}), as function of the dimensionless parameter $t_\bot \tau$ (in natural units). With respect to the experimental tight-binding parameters (\ref{experimentalpar}), the hopping $t_\bot$ sets the time scale $\tau_{\bot} = t_{\bot}^{-1} \sim 0.3 \, \mbox{eV}^{-1}$. For this plot, as well as for the following ones, it has been adopted $\Lambda/t_\bot = 1$ such that $\Lambda$ and $t_\bot$ have the same magnitude and are associated with the same timescale $\tau_{\bot}$\footnote{The general effects of the bias-voltage term on the LL entanglement of bilayer graphene was previously described in \cite{PRB}}. The quantum oscillation pattern exhibited by the survival probabilities has well-defined periodicities set by the characteristic periods
\begin{eqnarray}
\label{periodicites}
\tau_{C} = 2 \pi \,\left(\frac{\Lambda}{2}\right)^{-1},\qquad \mbox{and} \qquad
\tau_{W} =2 \pi \, \left( \Lambda^2+4 t_\bot^2 \right)^{-1/2},
\end{eqnarray} 
for Cat and the Werner states, respectively. The periods are defined by the differences between the eigenenergies from (\ref{eigenvalues}). For $\Lambda/t_{\bot} = 1$, they are related with the time scale $\tau_{\bot}$ by $\tau_C = 4 \pi \tau_\bot \sim 3.8 \, \mbox{eV}^{-1}$ and $\tau_W = 2 \pi\tau_{\bot}/\sqrt{5} \sim 2.8 \, \mbox{eV}^{-1}$. The oscillation amplitude associated to the Cat state is bound by the evolution from the initial configuration to its orthogonal state $\vert \psi_C ^- \rangle = ({\vert A1 (\bm{\bm{K}_+}) \rangle - \vert B2 (\bm{\bm{K}_+}) \rangle})/{\sqrt{2}}$. Otherwise, along the time evolution, the Werner state has a non-zero probability to be measured in its initial configuration. The above results show that the entanglement of the Cat state is unaffected by the time evolution while the entanglement of $\rho_W(\tau=0)$ oscillates, with upper bound plateau of maximum entanglement defined by the characteristic period, $\tau_W$.

\section{Noise effects on lattice-layer entanglement}

Once the free evolution of one-particle states is recovered by Eq.~(\ref{timeevo}), it is possible to include effects of classical noise into the dynamics. The action of the noise in a given quantum state is described through a time-dependent Hamiltonian, $\hat{\mathcal{H}}_{noise} (\tau)$. In the context of the Hamiltonian dynamics for bilayer graphene systems, as a first approach, it is assumed that the noise corresponds to random classical fluctuations of the bias voltage from Eq.~(\ref{biasvoltageHamiltonian}) as well as gap-opening fluctuations associated to the mass term from Eq.~(\ref{massHamiltonian}). The noise Hamiltonian is thus given by:
\begin{equation}
\label{Hnoise}
\hat{\mathcal{H}}_{noise} (\tau) = \frac{\Lambda^\prime (\tau)}{2} \, \hat{\sigma}_z^{(1)} \otimes \hat{I} + \frac{m^\prime(t)}{2} \, \hat{I} \otimes \hat{\sigma}_z^{(2)}.
\end{equation}
where lattice and layer DoF's are separately affected by the noise. In particular, it is assumed that $\Lambda^\prime (\tau)$ and $m^\prime(\tau)$ are modeled by the Ornstein-Uhlenbeck process characterized by the mean value properties \cite{OUProcess01, OUProcess02}
\begin{eqnarray}
\label{nonMarkov}
\langle \, A(\tau) \, \rangle = 0, \quad \langle \, A(\tau_i) \, A(\tau_j) \, \rangle = \frac{\Gamma_A \nu}{2} e^{- \nu \vert \, \tau_i - \tau_j \, \vert} \quad \quad (A \, = \, \Lambda^\prime, m^\prime ).
\end{eqnarray}
The fluctuations are non-Markovian with the correlation time defined by the noise bandwidth, $\nu$, and by a well-defined Markovian limit obtained as $\mbox{lim}_{\nu \rightarrow \infty} \, \langle A(\tau) \, A(s) \rangle = \Gamma_A \delta(\tau - s)$, that is, for infinite bandwidth or, conversely, for vanishing bath correlation time $T = \nu^{-1}$. Although non-Markovian dynamics imply into memory effects included via integrals of past times \cite{NM01}, under peculiar circumstances, it is possible to include the memory effects in time-dependent coefficients \cite{NM02, NM03, NM04}. Moreover, the inclusion of memory effects with time-dependent coefficients can be applied to descriptions of non-interacting qubits subjected to Ornstein-Uhlenbeck processes \cite{OUProcess02}.

As to recover the complete time evolution through the prescription from \cite{OUProcess02, YUNoise}, one writes the time evolution of a given initial state $\rho(0)$ in the interaction picture,
\begin{equation}
\label{evoeq}
\tilde{\rho} (\tau) = \exp \left[i \displaystyle \int_{0}^\tau \hat{H}_{noise} (s) ds \right] \rho(0) \exp \left[-i \displaystyle \int_{0}^\tau \hat{H}_{noise} (s) ds \right].
\end{equation}
The time evolved density matrix, $\tilde{\rho} (\tau)$, can be obtained as the solution of the master equation including the noise term which, for the process (\ref{nonMarkov}), reads \cite{NM03, NM04}
\begin{equation}
\frac{d \, \rho}{d \tau} = \frac{G(\tau)}{4} \bigg{(}2 \rho - \hat{I}\otimes \hat{\sigma}_z \rho \hat{I}\otimes \hat{\sigma}_z - \hat{\sigma}_z \otimes \hat{I} \rho \hat{\sigma}_z \otimes \hat{I} \bigg{)},
\end{equation}
assuming $\Gamma_{\Lambda^\prime} =\Gamma_{m^\prime} = \Gamma$, with $$G(\tau) = \int_{0} ^\tau \, ds \frac{\Gamma_A \nu}{2} e^{- \nu \vert \, \tau - s \, \vert} = \frac{\Gamma_A}{2}(1 - e^{- \nu \tau}).$$ The solution of the master equation can be written in a more compact form in terms of the Kraus operator sum representation \cite{Kraus}. By taking the statistical mean of (\ref{evoeq}) the behavior of $\tilde{\rho}(\tau)$ is given by \cite{OUProcess02}
\begin{equation}
\tilde{\rho} (\tau) = \displaystyle \sum_{\mu=1}^4 K_{\mu}^\dagger (\tau) \, \rho(0) K_\mu (\tau), 
\end{equation}
where $K_\mu$ are the Kraus operators associated to the noise, which are given by
\begin{eqnarray}
K_1(\tau) &=& E_1(\tau) \otimes E_1(\tau), \quad \quad K_2 = E_1(\tau) \otimes E_2 (\tau), \nonumber \\
K_3(\tau) &=& E_2(\tau) \otimes E_1(\tau), \quad \quad K_4 = E_2(\tau) \otimes E_2 (\tau),
\end{eqnarray}
where
\begin{eqnarray}
E_1(\tau)= \left[ \begin{array}{cc} p(\tau)& 0 \\ 0 & 1 \end{array}\right], \hspace{0.5 cm} E_2(\tau) = \left[ \begin{array}{cc} \sqrt{1 - p^2(\tau)} & 0 \\ 0 & 0 \end{array}\right],
\end{eqnarray}
and the time-dependent coefficient $p(\tau)$ is given in terms of $\Gamma$ and $\nu$ as
\begin{eqnarray}
 p(\tau)= \exp{[- f(\tau)]}, \hspace{0.5 cm} f(\tau) = \frac{\Gamma}{2} \left[ \tau + \frac{1}{\nu} (e^{-\nu \tau} - 1)\right].
\end{eqnarray}
The complete time evolution of the state in the Schr\"odinger picture can be recovered by using the completeness relation of the eigenstates (as in Eq.~(\ref{timeevo})) as to return
\begin{eqnarray}
\label{noiseevo}
\rho(\tau) &=& e^{i \mathcal{H} \tau} \tilde{\rho}(\tau) e^{- i \mathcal{H} \tau} \nonumber \\
&=& \displaystyle \sum_{n,s = 0}^1 \quad \sum_{m,l = 0}^1 \quad \sum_\mu e^{-i (\lambda_{n,s} - \lambda_{m,l})\tau } \rho_{n,s} K_{\mu}^\dagger (\tau) \, \rho(0) K_\mu (\tau) \rho_{m,l},
\end{eqnarray}
and, in particular, the survival probability of the state reads
\begin{eqnarray}
\label{survivprobnoise}
\mathcal{P}_{\rho(0)}(\tau) = \mbox{Tr}[\rho(0) \rho(\tau)] = \displaystyle \sum_{\mu = 1}^4 \, \sum_{n,s =0}^1 \, \sum_{m,l = 0}^1 e^{- i (\lambda_{n,s} - \lambda_{m,l}) \tau} \mbox{Tr}\Big[ \rho(0) \rho_{n,s} K_{\mu}^\dagger (\tau) \rho(0) K_\mu (\tau) \rho_{m,l} \Big].
\end{eqnarray}
The next step describes how the entanglement is affected by the non-Markovian noise and how the memory effects, encoded in the bandwidth $\nu$, influence the state dynamics.

\begin{figure}
\centering
\includegraphics[width= 14 cm]{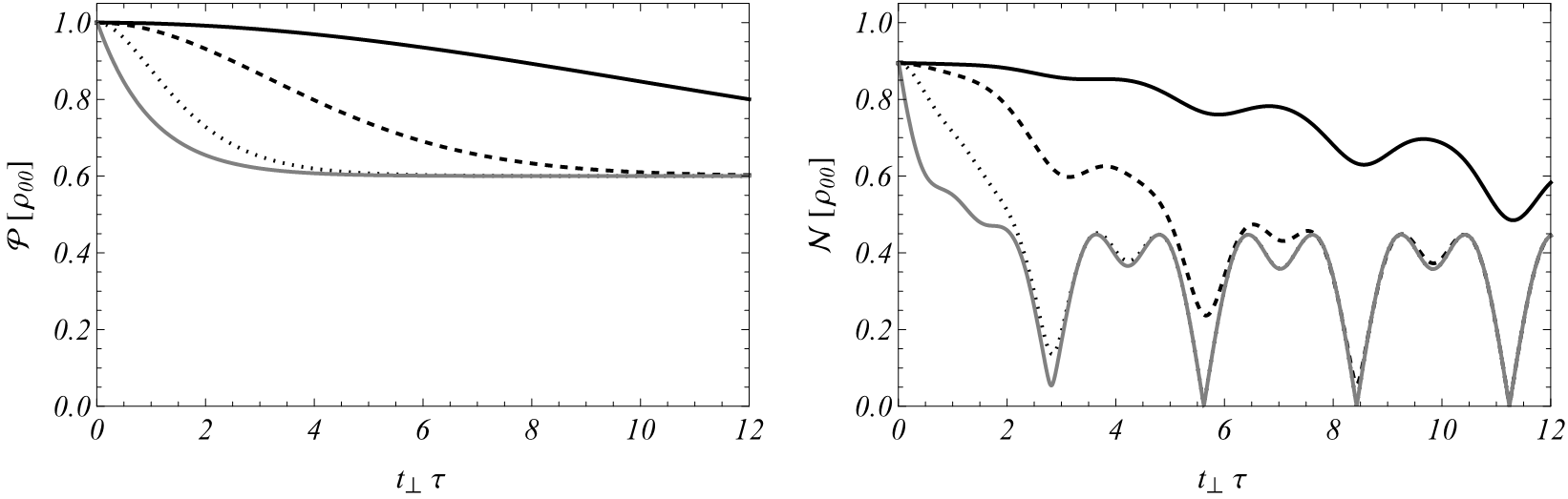}
\renewcommand{\baselinestretch}{1.0}
\caption{Survival probability (left plot) and negativity (right plot) for a state initially prepared as the positive energy eigenstate $\rho_{00}$ of the noiseless Hamiltonian (\ref{Htotal}) under the influence of the non-Markovian noise. The plots are for noise bandwidths $\nu/t_\bot = 0.01$ (thick line), $0.1$ (dashed line), $1$ (dotted line) and $\nu \rightarrow \infty$ (gray line), for the state with wave vector corresponding to the corner of the first Brillouin zone and with all other parameters in correspondence with Fig.~\ref{FigNoiselessSurvivor}. While the survival probability exhibits an exponential decay, the entanglement of the state tends to an oscillatory behavior (deaths and revivals). For $\tau \gg 1/t_\bot$, the entanglement does not depend on the noise bandwidth.}
\label{FigGridEig}
\end{figure}

Firstly, one considers the effects of the noise from Eq.~(\ref{Hnoise}) on the entanglement properties of a state prepared initially as an eigenstate of the noiseless Hamiltonian (\ref{ansatz}), with wave vector in the corner of the first Brillouin zone $\bm{K}_+$. Fig.~ \ref{FigGridEig} shows the survival probability and the negativity of a state initially prepared as the positive energy eigenstate $\rho_{00}$ as function of the parameter $t_\bot \, \tau$, for $\nu/t_{\bot} = 0.01$ (thick line), $0.1$ (dashed line), $1$ (dotted line) and in the Markovian limit $\nu/t_\bot \rightarrow \infty$ (gray line), thus capturing the effect of different orders of the environmental memory time scale. For example for $\nu/t_\bot = 0.01$, the environment memory scale, $\tau_{mem}$, is of the order of $\sim 1/\nu = 10^2 \, \tau_\bot$ which, for the experimental values of the hopping parameters (cf. Eq.~(\ref{experimentalpar})) is $\sim 38.1 \, \mbox{eV}^{-1}$. On the other hand, in the Markovian limit $\tau_\bot \gg \tau_{mem}$, and memory effects are related to time scales much smaller than the characteristic evolution scale set by the hopping parameter $t_\bot$. Assuming that $\Gamma/t_\bot = 1$ is equivalent to set that, in the Markovian limit, the noise will affect the state in the same time scale of the free evolution given by $\tau_{\bot}$. Additional parameters are in correspondence with those ones adopted in the noiseless case (cf. Fig.~\ref{FigNoiselessSurvivor}). 

\begin{figure}
\centering
\includegraphics[width= 14 cm]{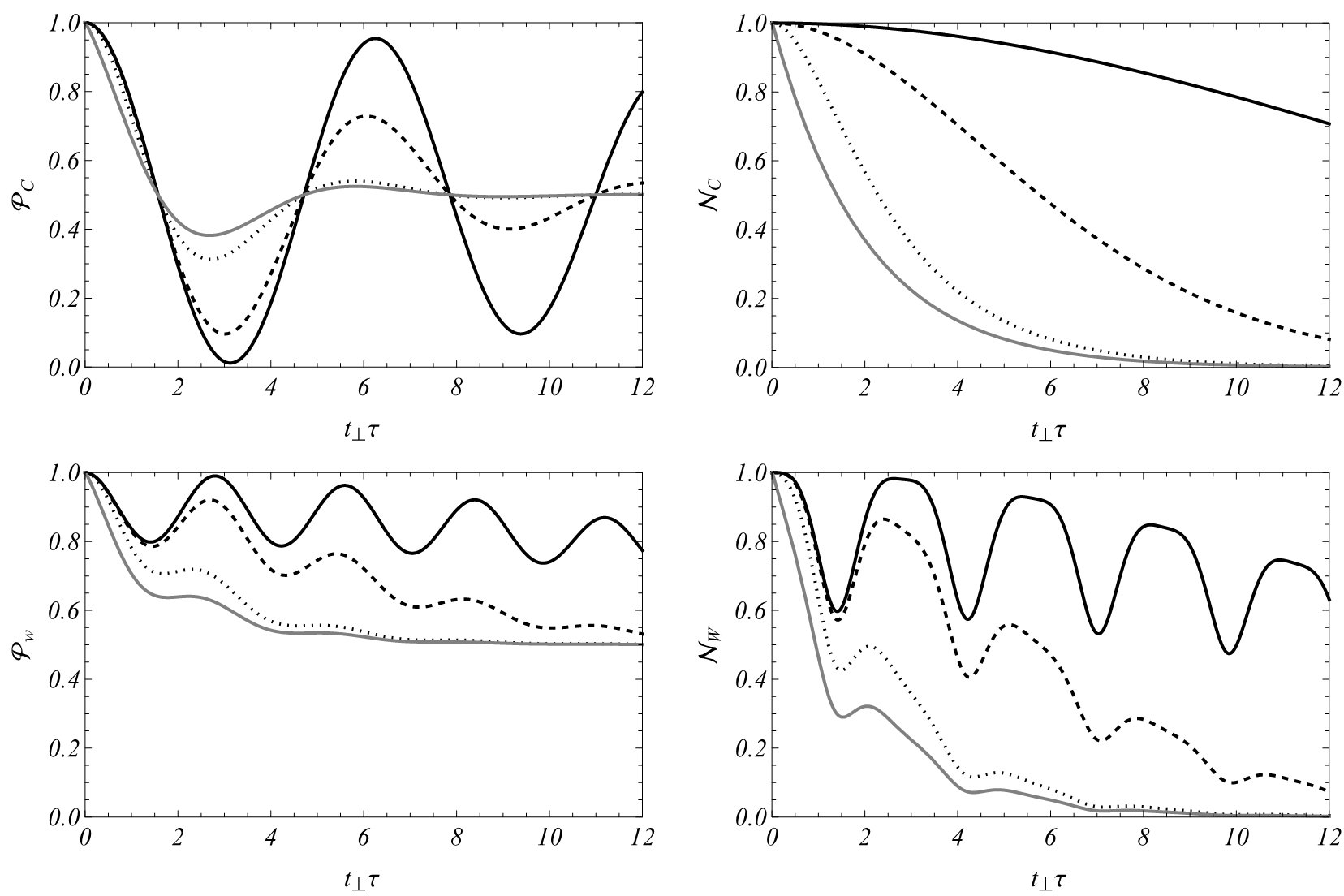}
\renewcommand{\baselinestretch}{1.0}
\caption{Survival probability (left column) and negativity (right column), for initial Cat (first row) and Werner (second row) states (\ref{cwstates}) under the non-Markovian noise. The parameters and plot styles are in correspondence with those of Fig.~\ref{FigGridEig}. The loss of entanglement due to the interaction with the environment has an exponential profile for the Cat state, and a non-monotonous damped oscillatory profile for the Werner state. In particular, small noise bandwidths (i.e. larger environment correlation times) usually preserve the initial characteristics of the states.}
\label{FigGridCW}
\end{figure}

The random fluctuations drive the state into a statistical mixture and the survival probability exhibits a monotonous decay. In the Markovian limit, the survival probability exhibits an exponential decay profile and quantum entanglement is also degraded due to the environment coupling. Nevertheless, the time-evolved state exhibits entanglement oscillations with death and revivals with defined frequency. For small noise bandwidths, the initial characteristics of the state are preserved for longer times and for $\tau \gg 1/t_\bot$ time-dependence of entanglement do not depend on the noise bandwidth. 
States initially set with maximal entanglement configurations (\ref{cwstates}) have the entanglement destroyed by the noise. Fig.~\ref{FigGridCW} shows the survival probabilities (left column) and the entanglement (right column) of initial Cat (first row) and Werner (second row) states.
Similar to the results depicted in Fig.~\ref{FigGridEig}, Cat and the Werner states have their initial configuration driven off by the noise.
Damped oscillations drive the system asymptotically to a statistical mixture with $50 \%$ of their original configuration. In both Cat and Werner cases, quantum entanglement is also degraded. The Cat state shows an exponential suppression profile of its initial entanglement, without oscillations, while the Werner state, even in the Markovian limit, has oscillations enveloped by the suppression rate. The non-Markovian term of the noise preserves the amount of entanglement, competing with the decoherence.
For larger interaction times with the environment, both states are completely disentangled, and as an eigenstate prospect, small noise bandwidths tends to preserve the initial characteristics of both states.

To end up, it is worth mentioning that additional relaxation processes, which might be relevant for describing transport properties of the bilayer graphene \cite{Relaxation01,Relaxation02}, can also affect the LL entangling properties. The main relaxation processes involved in the transport phenomena of bilayer graphene are related with electron-phonon and electron-electron scatterings, and impurities \cite{graph01, Relaxation01, Relaxation02}.
They all produce some energy loss of the material carriers \cite{Relaxation01, Relaxation02}. The electron-phonon interaction can be described by the inclusion of vector fields in the effective Dirac dynamics \cite{graph01}, which can lead, for instance, to localization effects on quantum states similar to those observed for the strained graphene . Electron-electron scatterings are included via Coulomb potentials in the tight-binding prescription, which demands for a more complex analysis. In both cases, a second quantization framework reveals some suitable transport properties \cite{graph01} from which, however, the corresponding description of a many-body influence on entanglement properties has not been worked-out. A challenging proposal could be related to the inclusion of electron and phonon interactions through open quantum system techniques, similar to those used in quantum optics and to describe ionic systems \cite{Breuer}, in a framework which also involve finite-temperature effects. In this case, the dynamics of an arbitrary initial state is given in terms of a master equation and electron and phonon heat baths would lead to the state thermalization which, in general, suppresses the quantum entanglement, although some other quantum correlations can persist \cite{Werlang}.  

Impurities and disorder effects \cite{graph01} are included in the TB model by means of short-range potentials in the Dirac equation \cite{Imp01, Imp02, Imp03, Imp04, Imp05}, and the corresponding scattering processes with the impurities can be considered to derive transport properties. The effect of disorders through short ranged potentials can be evaluated by spherical wave scatterings in a framework similar to that used for computing the spin-parity entanglement under a barrier scattering \cite{Barrier}. In this case, the role of localization aspects should also be investigated.

\section{Conclusions}

In this work, the relation between the TB formulation of graphene interactions and the intrinsic entangled structure of Dirac equation solutions has been translated into a self-consistent formulation of the LL entanglement of single particle excitations of bilayer graphene.
Once the noiseless dynamics of an arbitrary initial states is recovered through the relation between the TB Hamiltonian and the modified Dirac Hamiltonian, the effects of a noise environment, modeled by Ornstein-Uhlenbeck process, through the Kraus operator sum representation, have been considered in order to suggest more realistic setups involving the LL entanglement.

The noise model considered here describes random fluctuations of bias voltage and mass terms (related to gap opening between the electronic bands of the system) and has a well-defined Markovian limit, which has allowed for investigating the noise memory effects on LL entanglement. 
For a state initially prepared as an eigenstate of the noiseless Hamiltonian, the survival probability shows an exponential decay profile under noise effects even whether, for long time interactions, the entanglement tends to an oscillatory behavior with death and revivals at definite frequencies. 

When Cat and Werner states are considered from the beginning, the initial entanglement is completely degraded by the noise environment and the states evolve into separable mixed states. 
While the Cat state entanglement exhibits an exponential suppression, the Werner state entanglement shows some non-monotonous decay.
In both cases, low noise bandwidths, associated with highly non-Markovian effects, in general, preserve the initial characteristics of a given state, and the Markovian limit is associated to a faster decoherence effect.

Our results follows the Hamiltonian dynamics description that have already supported some engendered Dirac-like configurations of non-relativistic physical systems \cite{DiracEmu01, DiracEmu02, DiracEmu03, DiracEmu04, DiracEmu05}. For example, for mapped Jaynes-Cummings Hamiltonians associated to trapped ions setups, an analogous Dirac dynamics including external fields have been constructed as to reproduce controllable relativistic-like effects \cite{DiracEmu01, Trapped01}. With the Dirac equation solutions reinterpreted in terms of ionic variables \cite{Trapped03}, the spin-parity entanglement is translated into the entanglement between total angular momentum and its projection onto the magnetic field responsible for lifting the ionic energy levels \cite{PRA}.
As performed in this paper, the framework including global noise effects that couple both DoF's of the system \cite{YUNoise, MyNoise} has been encompassed by the Dirac dynamics as to provide the setup for including random fluctuations of physically relevant parameters associated the quantum dynamics of the system. In such a context, still in the open quantum system formalism, other environment effects, such as coupling with a bosonic bath, can be described via a proper master equation whose solutions often require numerical techniques \cite{Breuer}.

As a last remark, although no protocol for direct single-particle state manipulation in graphene is available, the increasing of the number of protocols on graphene experimental characterization possibly supports the building of quantum gates using the qubit assignment (\ref{assi}) discussed here.
To construct protocols to map the entanglement encoded in the internal DoF's of a single particle into entanglement between the DoF's of two particles \cite{Intra01}, the systematic characterization of open quantum system effects in the qubit state is relevant for devising error-correction methods as well as for characterizing the engineering of quantum gates.
The construction of quantum gates with operation time shorter than the system intrinsic decoherence time \cite{Breuer, Chuang} involving the characterization quantum correlations in mixed states of the bilayer graphene deserve further investigations and are all postponed to future issues.

{\em Acknowledgments} - The work of AEB is supported by the Brazilian Agencies FAPESP (grant 2017/02294-2) and CNPq (grant 300831/2016-1). The work of VASVB is supported by the Brazilian Agency CAPES (grant 88881.132389/2016-1).

\section*{Appendix -- AB stacking scheme}
\begin{figure}[H]
\centering
\includegraphics[width= 12 cm]{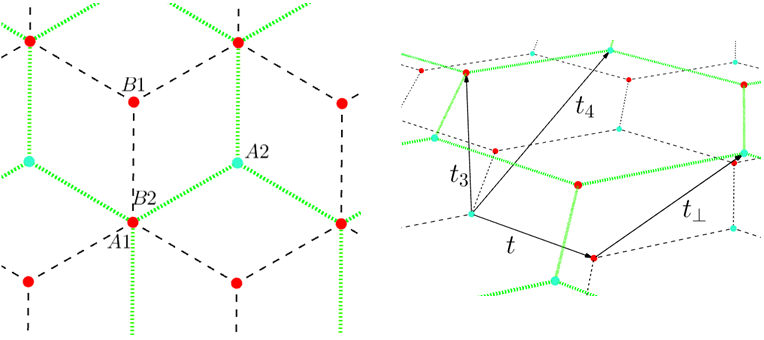}
\renewcommand{\baselinestretch}{1.0}
\caption{ 
Top view of the geometry of the AB (Bernal) stacking (left) and schematic representation of the hopping amplitudes of Eq.~(\ref{tightbinding}) (right) -- scheme reproduced from Ref.~\cite{PRB}.}
\label{GeoABStack}
\end{figure}

In the scheme Fig.~\ref{GeoABStack} reproduced from Ref.~\cite{PRB} for the AB stacking \cite{graph03, graph04,PRB}, half of the atoms of the upper layer (joined by dotted lines) are localized exactly above half of the atoms of the lower layer (joined by dashed lines). Sites that are placed exactly above a site of the lower layer are called dimer sites (A1 and B2), while sites that are localized above the center of the other honeycomb are called non-dimer sites (B1 and A2). The hopping amplitudes of the TB model for the bilayer graphene in AB stacking are: $t$ describing the hopping between next neighbors in the same layer; $t_\bot$ describing the hopping from a non-dimer site to its nearest non-dimer site; $t_3$ describing the hopping from a dimer site to its nearest dimer site, and finally, $t_4$ describing the hopping from a dimer to the nearest non-dimer site. In each layer, the lattice is formed by two superposed sublattices, labeled by $A$ and $B$.
The presence of the interlayer hopping $t_3$ produces distortions onto the iso-energy lines around the Dirac points -- the trigonal wrapping -- and, for large values of $t_3/t_\bot$, additional local minimum energy points are evinced \cite{graph03, graph04, Predin}. The effects of such interlayer coupling can also be observed in the entanglement spectrum of single particle excitations \cite{Predin,PRB}, as well as in conductivity \cite{Trig01} and interference effects \cite{Trig02}.

\end{document}